\documentclass{webofc}
\usepackage[varg]{txfonts}   
\usepackage[colorlinks=true,linktocpage=true,linkcolor=blue,citecolor=blue,allcolors=blue]{hyperref}
\begin{document}
\title{Fluctuations of conserved charges in hydrodynamics and molecular dynamics}
%
%

\author{\firstname{Volodymyr} \lastname{Vovchenko}\inst{1,2,3}\fnsep\thanks{\email{vlvovch@uw.edu}} 
}

\institute{Institute for Nuclear Theory, University of Washington, Seattle, WA 98195, USA
\and
           Frankfurt Institute for Advanced Studies, D-60438 Frankfurt am Main, Germany
\and
           Nuclear Science Division, Lawrence Berkeley National Laboratory, Berkeley, CA 94720, USA 
          }

\abstract{%
I present an overview of recent theoretical results on fluctuations of conserved charges in heavy-ion collisions obtained in relativistic hydrodynamics and molecular dynamics frameworks. In particular, I discuss the constraints on the location of the QCD critical point based on comparisons of experimental data on proton number cumulants with precision calculations of non-critical contributions. Recent developments on critical fluctuations in molecular dynamics simulations are covered as well. 
}
\maketitle
\section{Introduction}
\label{intro}

Fluctuations of conserved charges are sensitive probes of QCD matter.
In thermodynamic equilibrium within the grand-canonical ensemble, the cumulants of a conserved charge distribution are linked to the susceptibilities, i.e., to the chemical potential derivatives of the partition function, 
\begin{equation}
\kappa_n \propto \frac{\partial^n (\ln Z^{\rm gce})}{\partial \mu^n}.
\end{equation}
Therefore, fluctuations probe the finer details of the equation of state.
In particular, the conserved baryon number plays the role of the order parameter for the hypothetical first-order QCD phase transition at finite baryon densities, and its fluctuations exhibit singular behavior near the critical point~(CP) of the transition~\cite{Stephanov:2004wx, Stephanov:2008qz}. 
Critical opalescence is a classical example of such a phenomenon when a usually transparent substance with respect to laser irradiation becomes opaque due to large density fluctuations of macroscopic scales near the CP.
The location~(and even existence) of the QCD CP and the associated phase structure of QCD matter are one of the most important open questions tackled by relativistic heavy-ion collisions at various energies~\cite{Bzdak:2019pkr}. 

In contrast to classical fluids, it is impossible to trap a droplet of hot and dense QCD fluid to try to observe critical opalescence.
The blob of QCD matter created in heavy-ion collisions quickly expands and hadronises, producing at most a few thousands of (anti)baryons and typically even less.
On the other hand, the relative ``smallness'' of the number of particles created in heavy-ion collisions makes it possible to track each particle in each event~(modulo detector efficiency and acceptance limitations), and compute the event-by-event distribution and the associated fluctuation measures directly.
In this regard, the cumulants of the proton number -- a proxy for the baryon number in the experiment -- are sensitive probes of critical behavior~\cite{Hatta:2003wn}, in particular the high-order non-Gaussian cumulants~\cite{Stephanov:2008qz, Stephanov:2011pb}.
The cumulants of proton number are also used to extract other information, and recent works have studied, for instance, the speed of sound~\cite{Sorensen:2021zme} or freeze-out temperature~\cite{Kitazawa:2022gmq}.

Measurements of proton number fluctuations are being performed by several experiments, including ALICE~\cite{ALICE:2019nbs}, STAR~\cite{STAR:2020tga,STAR:2021iop,STAR:2021fge}, NA61/SHINE~\cite{Gazdzicki:2017zrq}, and HADES~\cite{HADES:2020wpc}.
Much attention is devoted to measuring the net-proton kurtosis $\kappa \sigma^2$ at RHIC beam energy scan, which indicated a possible non-monotonic collision energy dependence of this quantity~\cite{STAR:2020tga}. 
Such a feature has earlier been predicted as a potential signal of the QCD critical point~\cite{Stephanov:1999zu}.
It should be noted, however, that experimental uncertainties are significant, and some of the observed features could be attributed to baryon conservation as opposed to critical point~\cite{Braun-Munzinger:2020jbk}.
More robust conclusions will become possible once the improved data on $\kappa \sigma^2$ from BES-II becomes available.
In the meantime, however, one could explore what can be learned from the more precise available measurements of the second and third-order proton cumulants.
These analyses, though, require quantitative comparisons between theory and experiment, which for event-by-event fluctuations involve many caveats~(see e.g.~\cite{Vovchenko:2021gas} for an overview). 
Addressing these caveats requires extensive dynamical modeling of heavy-ion collisions.

Several strategies are being considered to tackle the search for the QCD critical point in heavy-ion collisions within dynamical models:
\begin{itemize}
    \item \textit{Precision calculation of non-critical proton number fluctuations and its deviations from experimental data.}
    
    This approach, recently developed in Ref.~\cite{Vovchenko:2021kxx}, incorporates essential non-critical contributions to proton number fluctuations~(conservation laws, hadronic interactions, momentum cuts) on top of the hydrodynamical background. Recent results are summarized in Sec.~\ref{sec:hyd}.
    
    \item \textit{Molecular dynamics with a critical point.}
    
    \emph{Molecular dynamics} is a microscopic approach that can address many of the caveats associated with fluctuations. Insights on critical fluctuations using classical molecular dynamics have recently been explored in Ref.~\cite{Kuznietsov:2022pcn} and covered in Sec.~\ref{sec:MD}.
    
    \item \textit{Hydrodynamics with critical fluctuations.}
    
    Ultimately, critical fluctuations should be incorporated into the hydrodynamic framework for heavy-ion collisions~\cite{Stephanov:2017ghc} as well the particlization procedure~\cite{Pradeep:2022mkf}, allowing one to make testable predictions based on the location of the CP. Such a framework has been under development, for instance, within the BEST Topical Collaboration~\cite{An:2021wof}.
    
\end{itemize}

\section{Hydrodynamics based analysis of (net-)proton fluctuations}
\label{sec:hyd}

One way to search for critical behavior in proton number cumulants is to study deviations of experimental data from baseline predictions that do not incorporate critical fluctuations.
The simplest baseline corresponds to uncorrelated proton production, which would yield proton number cumulants consistent with Poisson statistics.
However, additional non-critical mechanisms such as baryon number conservation and the repulsive core in baryon-baryon interaction break this assumption and make the non-critical baseline considerably more involved.
Relativistic hydrodynamics is considered to be the standard model of heavy-ion collisions and provides a realistic background for calculating the aforementioned non-critical contributions to proton number cumulants.
More specifically, the effects of baryon conservation and repulsion are implemented at the Cooper-Frye particlization stage, which is performed either analytically~\cite{Vovchenko:2021kxx, Vovchenko:2021yen} or through Monte Carlo sampling~\cite{Vovchenko:2022syc}. Baryon repulsion is modeled by utilizing the excluded volume model, with baryon excluded volume parameter $b = 1$~fm$^3$ as follows from fits to lattice QCD susceptibilities~\cite{Karthein:2021cmb}.

{\bf LHC.}
Net-proton number cumulants have been measured up to third order by the ALICE Collaboration at the LHC~\cite{ALICE:2019nbs, ALICE:2022xpf}.
Although fluctuations at the LHC are not expected to be sensitive to the possible presence of the QCD critical point in the baryon-rich regime, the available measurements of a normalized net-proton number variance $\kappa_2[p-\bar{p}]/\langle p+\bar{p} \rangle$ appear to establish the relevance of non-critical effects, in particular, that of baryon number conservation~\cite{ALICE:2019nbs, Vovchenko:2020kwg}.
Recently, the authors of Ref.~\cite{Savchuk:2021aog} pointed out that precision measurements of $\kappa_2[p-\bar{p}]/\langle p+\bar{p} \rangle$ at the LHC are also sensitive both to the range of baryon conservation as well as to baryon annihilation in the hadronic phase.
Namely, a surplus of annihilation over regeneration leads to an increase of $\kappa_2[p-\bar{p}]/\langle p+\bar{p} \rangle$ while reducing the range of correlations associated with baryon conservation brings this quantity down.
The data can be described equally well by employing either (1) global baryon conservation and no baryon annihilation or (2) local baryon conservation across $\Delta y_{\rm cor} \sim 3$ units of rapidity and baryon annihilation without regeneration modeled by UrQMD afterburner.
The two effects can be constrained experimentally by a combined measurement of $\kappa_2[p-\bar{p}]/\langle p+\bar{p} \rangle$ and  $\kappa_2[p+\bar{p}]/\langle p+\bar{p} \rangle$, while additional constrains can come from fluctuation measurements involving light nuclei~\cite{ALICE:2022amd}.

\begin{figure}[h]
\centering
\includegraphics[width=.45\textwidth,clip]{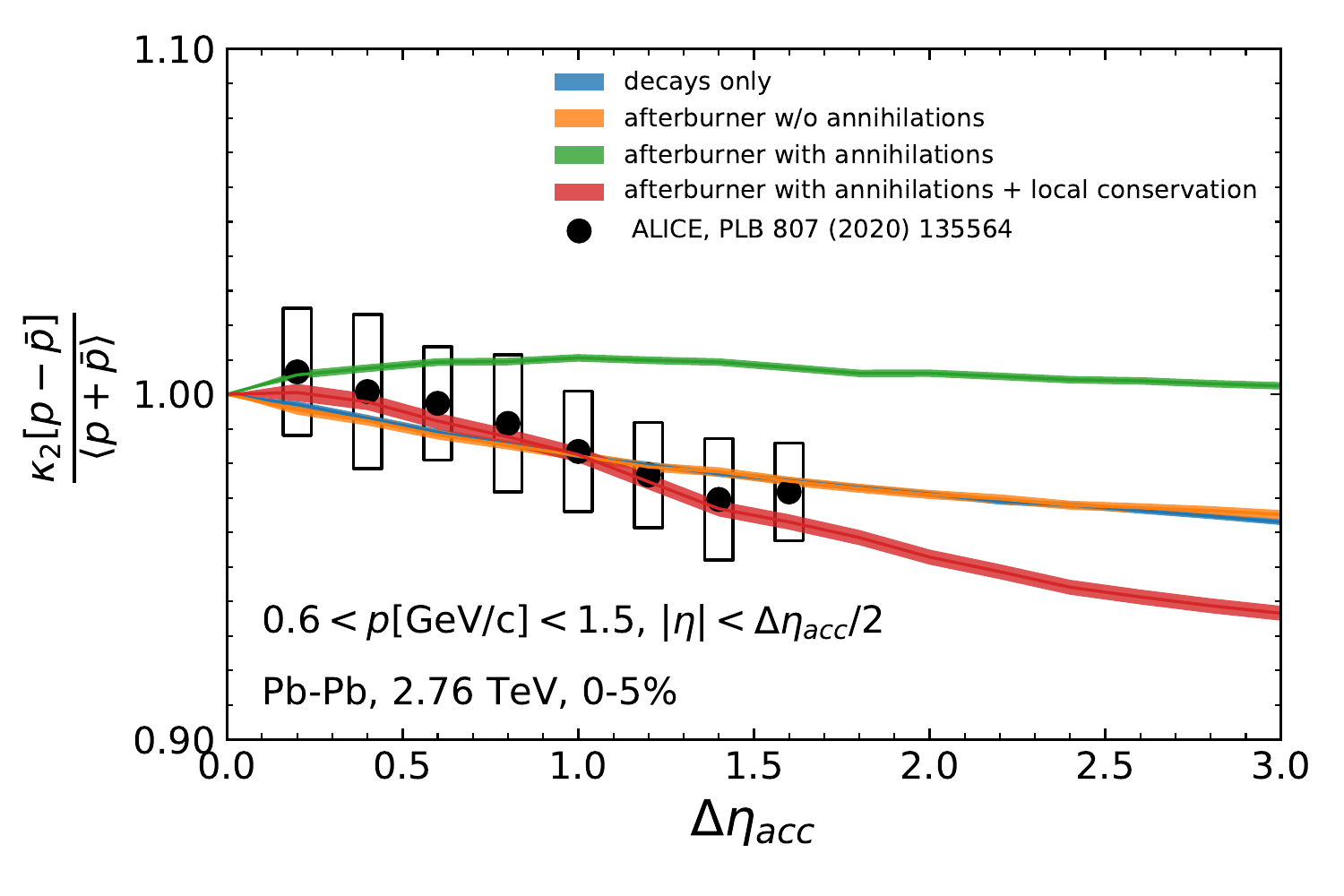}
\includegraphics[width=.45\textwidth,clip]{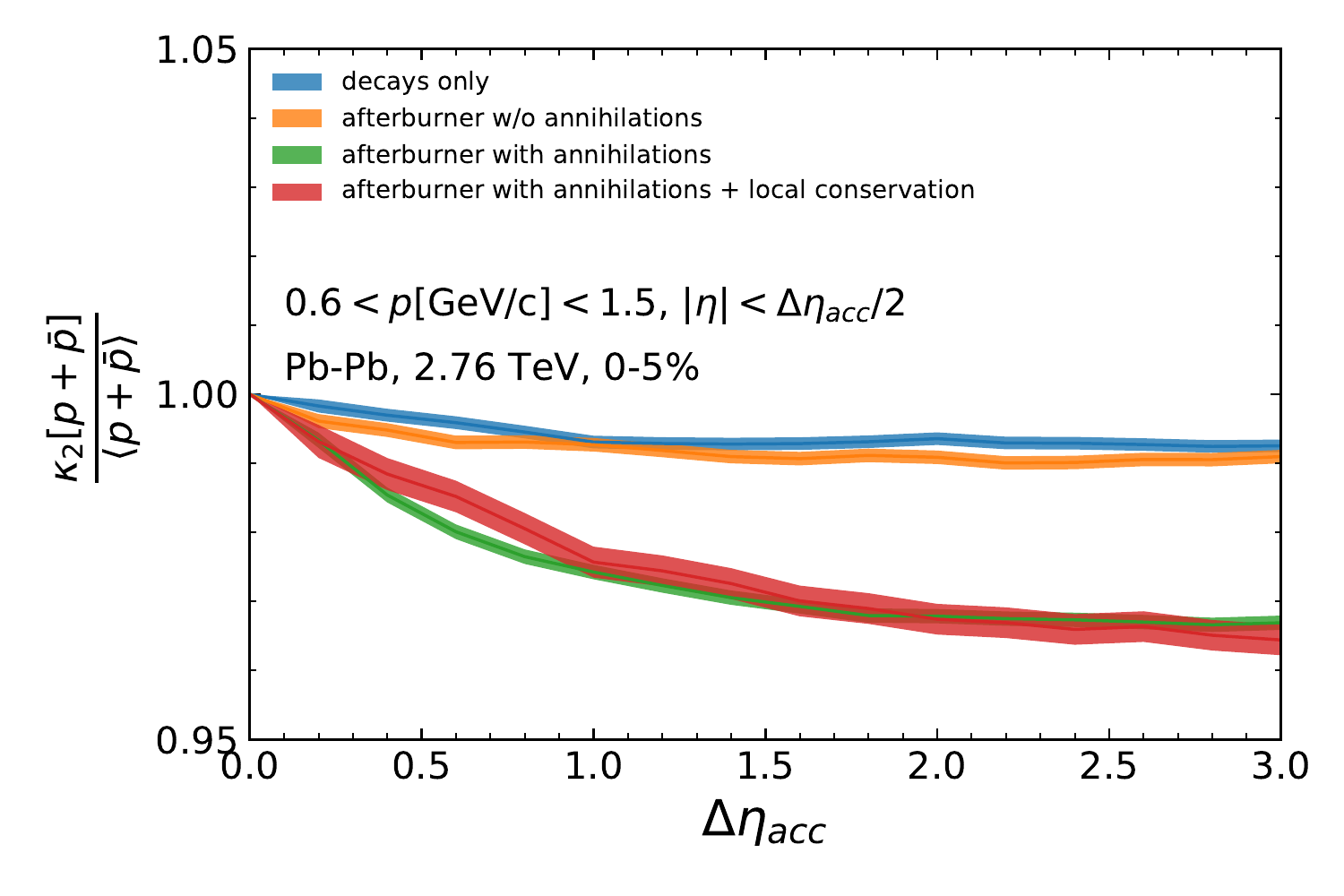}
\caption{
Dependence of proton cumulant ratios $\kappa_2[p-\bar{p}]/\langle p+\bar{p} \rangle$~(left) and $\kappa_2[p+\bar{p}]/\langle p+\bar{p} \rangle$~(right) on the pseudorapidity acceptance in 2.76 TeV Pb-Pb collisions calculated in different scenarios regarding the treatment of baryon annihilation and the range of baryon conservation.
Taken from~\cite{Savchuk:2021aog}.
}
\label{fig:LHC}       
\end{figure}

{\bf RHIC-BES.}
Proton number cumulants at RHIC-BES energies~($\sqrt{s_{\rm NN}} = 7.7-200$~GeV) have been measured by the STAR Collaboration~\cite{STAR:2020tga,STAR:2021iop}.
The non-critical contributions were calculated in Ref.~\cite{Vovchenko:2021kxx} based on hydrodynamic simulations of 0-5\% central Au-Au collisions within the MUSIC code~\cite{Shen:2020jwv}, where both the ordinary as well as factorial cumulants~\cite{Bzdak:2016sxg} of (net-)proton number distribution were analyzed.
It was shown that multi-proton~($n>3$) correlations are small in the non-critical scenario. Thus, the behavior of all the cumulants is driven by two-proton correlations, which are found to be negative.
This behavior contrasts critical fluctuations, which would generate sizeable multi-particle correlations~\cite{Ling:2015yau}.

\begin{figure}[h]
\centering
\includegraphics[width=.44\textwidth,clip]{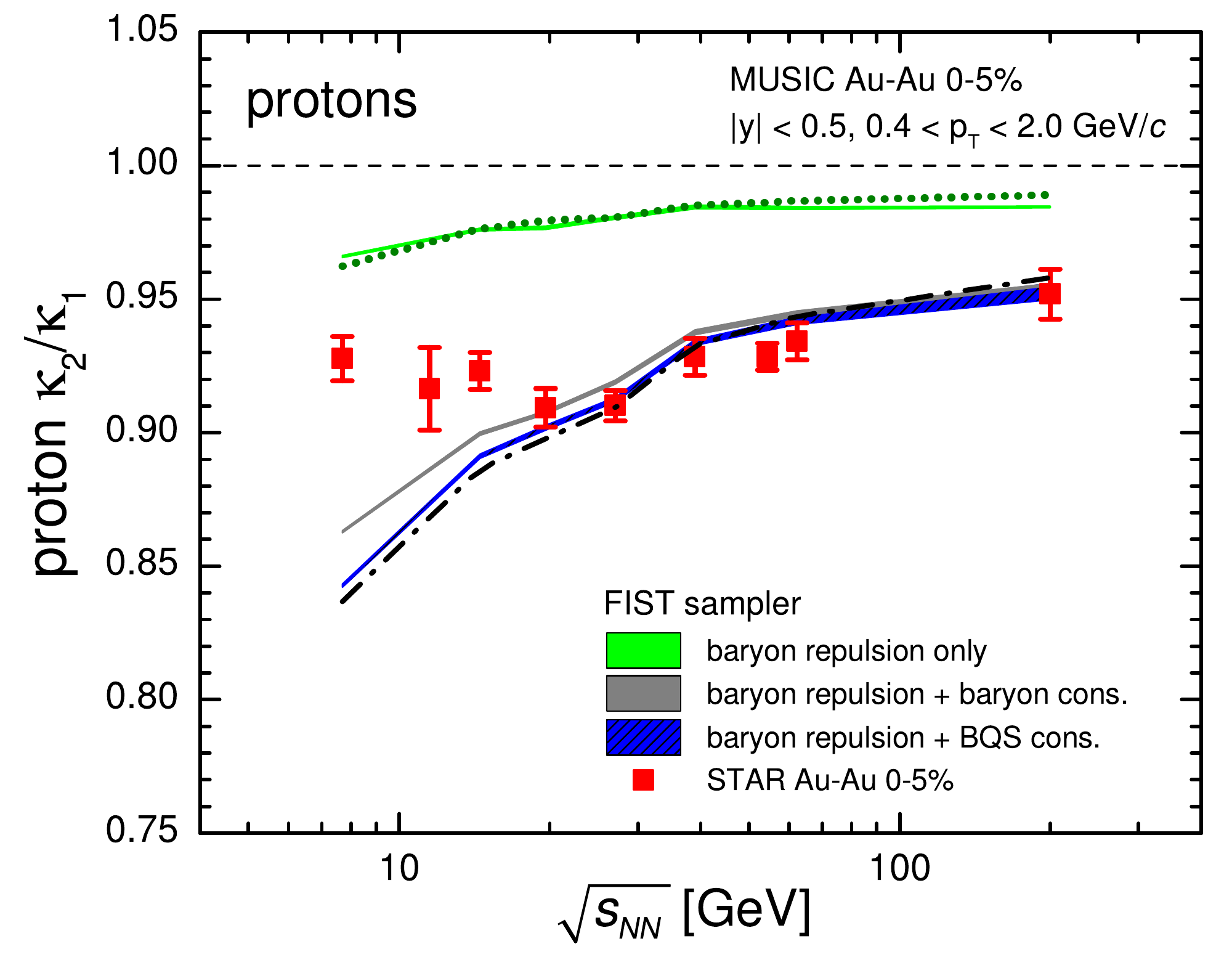}
\hskip3pt
\includegraphics[width=.49\textwidth,clip]{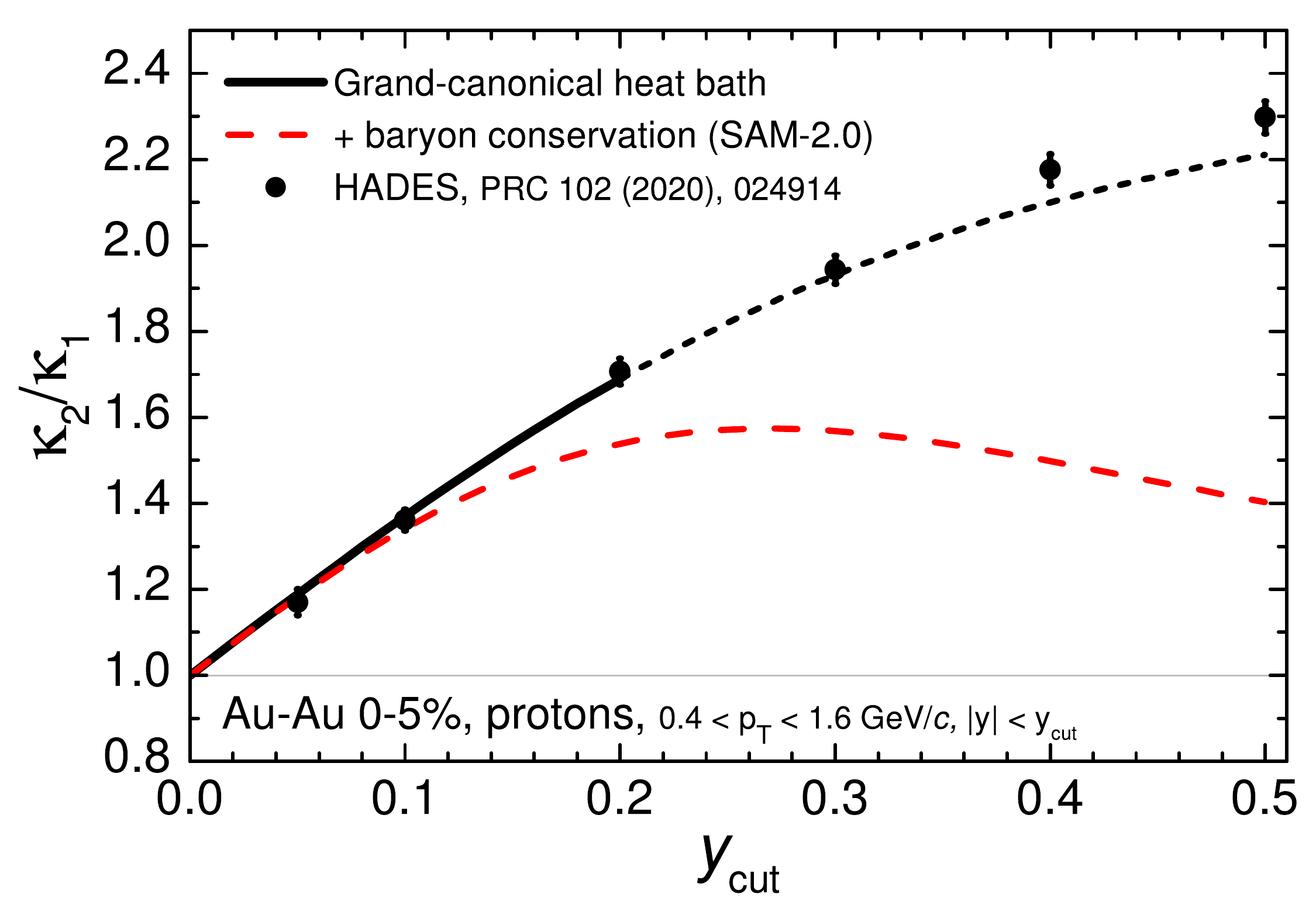}
\caption{
Scaled variance $\kappa_2/\kappa_1$ of the proton number distribution in central Au-Au collisions as a function of (left panel) collision energy and (right panel) acceptance in rapidity at $\sqrt{s_{\rm NN}} = 2.4$~GeV, incorporating various non-critical contributions.
Adapted from~\cite{Vovchenko:2022szk, Vovchenko:2022syc}.
}
\label{fig:BES}       
\end{figure}

The experimental data of the STAR Collaboration is quantitatively consistent with simultaneous effects of baryon conservation and repulsion at $\sqrt{s_{\rm NN}} \gtrsim 20$~GeV~(Fig.~\ref{fig:BES}), i.e. with non-critical physics.
At lower collision energies, however, the data indicate an excess proton correlation over the non-critical baseline.
This excess becomes even more prominent when new data from STAR fixed target programme at $\sqrt{s_{\rm NN}} = 3$~GeV~\cite{STAR:2021fge} and HADES experiment at GSI at $\sqrt{s_{\rm NN}} = 2.4$~GeV~\cite{HADES:2020wpc} is considered.

Additional non-critical contributions such as volume fluctuations and electric charge conservation can influence proton number cumulants.
In particular, adding volume fluctuations via an additional parameter can improve the data description at lower collision energies, but it will spoil the agreement at higher collision energies.
The effect of exact conservation of electric charge was explored in Ref.~\cite{Vovchenko:2022syc} and did not yield any improvement in the description of the data at $\sqrt{s_{\rm NN}} \lesssim 20$~GeV.

{\bf Lower energies.}
One can conclude that it is challenging to understand the data at $\sqrt{s_{\rm NN}} \lesssim 20$~GeV in terms of non-critical physics.
In particular, the HADES data point shows a large proton scaled variance of $\kappa_2 / \kappa_1 > 2$, warranting a closer look at these data. 
Proton number cumulants in Au-Au collisions at $\sqrt{s_{\rm NN}} = 2.4$~GeV have recently been analysed in \cite{Vovchenko:2022szk} in the framework of Siemens-Rasmussen-like fireball model, with freeze-out parameters based on Refs.~\cite{Harabasz:2020sei,Motornenko:2021nds}.
In contrast to the non-critical baseline calculation, this analysis requires no assumptions on baryon number susceptibilities characterizing the emitting source. Instead, one extracts their values from the data by fitting the first four proton cumulants at different rapidity acceptance cuts. In order to minimize the effects of baryon conservation, the cuts were considered up to $y_{\rm cut} \leq 0.2$.

The experimental data of HADES Collaboration is described by assuming a thermal emission of nucleons from a grand-canonical heat bath~(Fig.~\ref{fig:BES}), provided that the corresponding baryon number susceptibilities of QCD matter are highly non-Gaussian and exhibit the following hierarchy:
$\chi_4^B \gg -\chi_3^B \gg \chi_2^B \gg \chi_1^B$.
This kind of hierarchy of conserved charge susceptibilities can appear in the vicinity of a critical point~\cite{Stephanov:2008qz, Vovchenko:2015pya}.
Therefore, naively, this observation could point to a presence of the QCD critical point close to the HADES chemical freeze-out at $T \sim 70$~MeV and $\mu_B \sim 850-900$~MeV.

However, the behavior of proton cumulants at $y_{\rm cut} > 0.2$ is challenging to describe even qualitatively when one incorporates the effect of exact baryon conservation into the calculations~(dashed red line).
The above statement applies for $\kappa_2$ shown in Fig.~\ref{fig:BES} as well as for $\kappa_3$ and $\kappa_4$ not shown here,
indicating that more theoretical and experimental effort is required to reach a firm conclusion.
We emphasize that the challenge in understanding the results in the context of baryon conservation persists already at the second-order, $\kappa_2/\kappa_1$, and should be resolved at this level before turning to third- and fourth-order cumulants.

\section{Critical point particle number fluctuations from molecular dynamics}
\label{sec:MD}

Molecular dynamics~(MD) is a microscopic approach to studying (non-)equilibrium evolution of dynamical systems and provides an alternative approach to model heavy-ion collisions.
Most MD codes such as UrQMD or SMASH are purely hadronic, given that it is challenging to implement both hadron and quark degrees of freedom.
Nevertheless, such an approach is better suited to describe non-equilibrium evolution and is arguably preferable for intermediate collision energies.
MD simulations already provide useful insights into various non-critical contributions to particle number fluctuations~\cite{Nahrgang:2009dqc,STAR:2021fge,Hammelmann:2022yso}.

Critical fluctuations can also be studied in MD simulations, as long as one uses an appropriate interaction potential that leads to the presence of a first-order phase transition and a CP.
One famous and well-studied example of such a potential is the Lennard-Jones~(LJ) potential, which describes fluids exhibiting a liquid-gas type transition.
The LJ fluid corresponds to a system of non-relativistic particles with attractive and repulsive interactions.
This system is quite different from QCD matter, as QCD contains different degrees of freedom -- hadron and partons -- on the different sides of the QCD transition. 
Nevertheless, due to the universality of critical behavior, simulations of the LJ fluid can provide helpful insight into the microscopic development of critical fluctuations.

Recent work~\cite{Kuznietsov:2022pcn} used MD simulations of the LJ fluid to study the behavior of particle number fluctuations near and away from the CP through a numerical solution of Newton's equations of motion~(classical $N$-body problem).
Simulations were performed in a box with periodic boundary conditions and within the microcanonical ensemble.
Since the total particle number $N$ is conserved, fluctuations were studied in subsystems by performing cuts either in longitudinal coordinate $z$ or longitudinal velocity $v_z$.

\begin{figure}[h]
\centering
\includegraphics[width=.47\textwidth,clip]{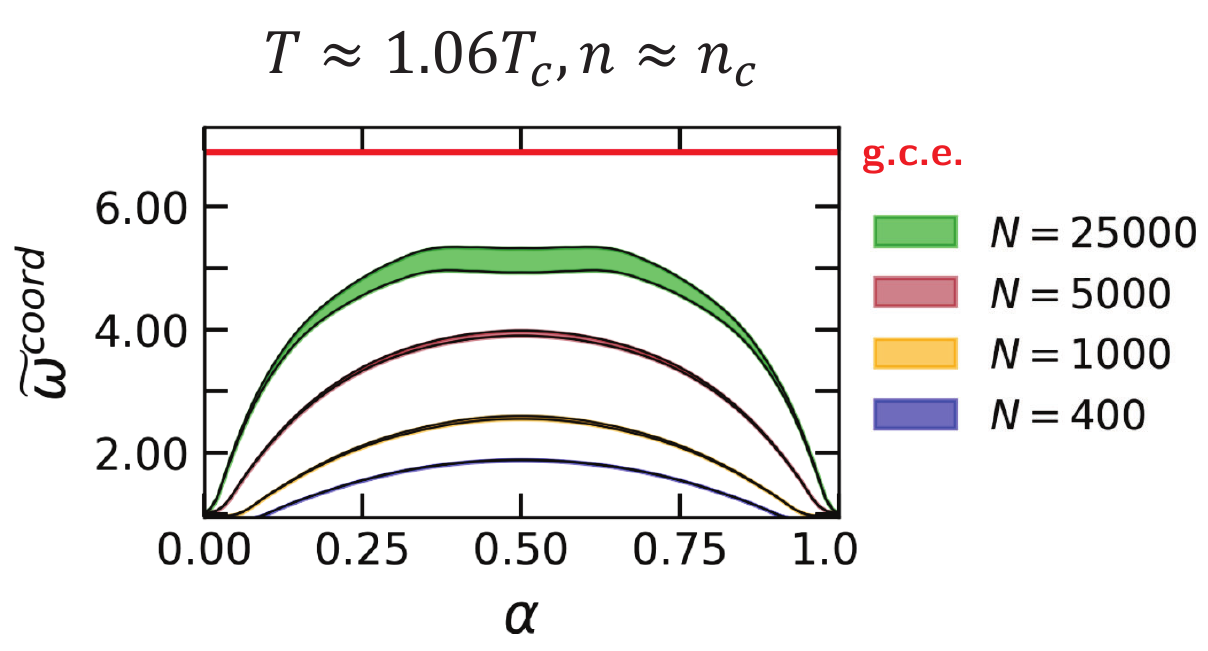}
\hskip15pt
\includegraphics[width=.36\textwidth,clip]{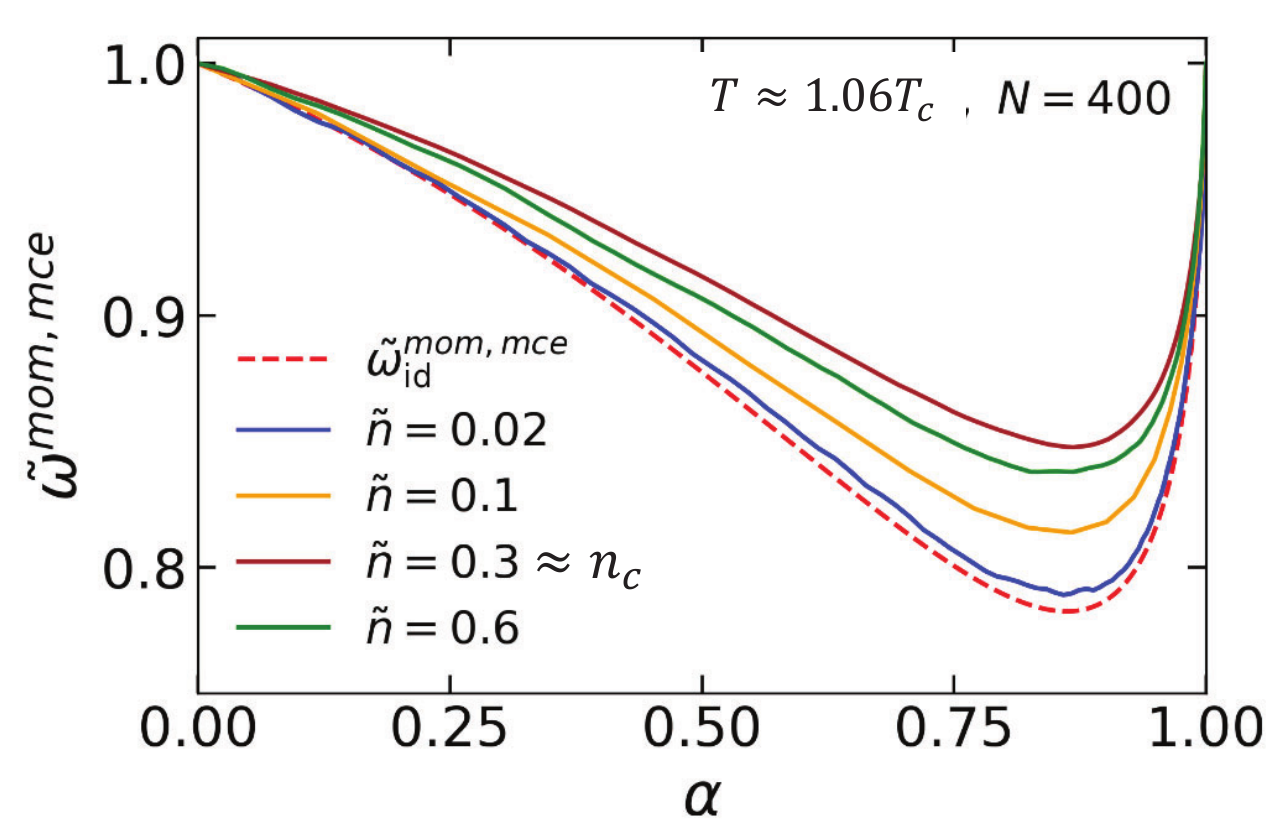}
\caption{
Dependence of the scaled variance of particle number corrected for global particle number conservation on the subvolume fraction $\alpha$ in coordinate~(left panel) and momentum~(right panel) space calculated using molecular dynamics simulations of the Lennard-Jones fluid near the critical point. 
Adapted from~\cite{Kuznietsov:2022pcn}.
}
\label{fig:MD}       
\end{figure}

The left panel of Fig.~\ref{fig:MD} depicts the results for the scaled variance $\tilde{\omega} = \omega/(1-\alpha)$ of particle number fluctuations corrected for total particle number conservation by $1-\alpha$ factor~\cite{Vovchenko:2020tsr}, as function of acceptance fraction $\alpha$ along the $z$-coordinate. The calculations were performed near the critical point~($T \simeq 1.06 T_c$, $n \simeq n_c$), and for different values of the total particle number $N$.
The results are consistent with an approach toward the grand-canonical scaled variance of $\omega^{\rm gce} \sim 7$ as the particle number~(system size) is increased. 
These observations indicate that the presence of the CP manifests itself in large particle number fluctuations and finite system-size effects, as expected.

However, when one studies the fluctuations in the momentum subspace instead, via a cut $|v_z| < v_z^{\rm cut}$ in longitudinal velocity~(right panel of Fig.~\ref{fig:MD}), the effect of critical fluctuations is completely washed out.
This observation reflects that coordinates and momenta in uniform LJ fluid are uncorrelated. Thus, the significant correlations in coordinate space due to CP do not translate into the momentum space.
The situation may be different in expanding systems encountered in heavy-ion collisions, where collective velocities generate correlations between coordinates and momenta of emitted particles and may preserve CP signals even in momentum space which is accessible experimentally.
Extensions of MD simulations with critical fluctuations to expanding systems will be the subject of future works.

\section{Summary}

\begin{figure}[h]
\centering
\hskip-40pt
\includegraphics[width=.65\textwidth,clip]{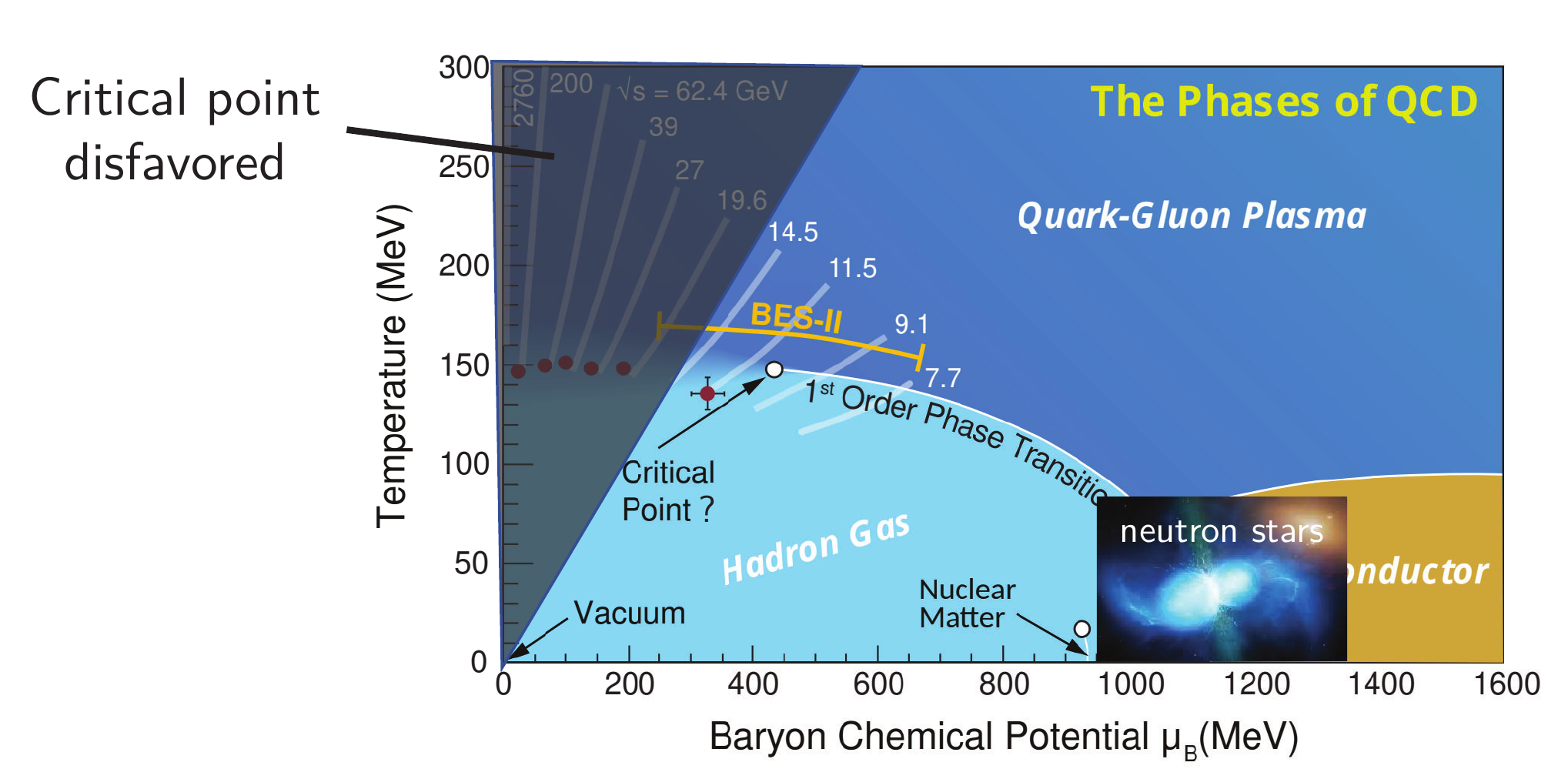}
\caption{
The figure shows the disfavored region for the QCP CP location on the QCD phase diagram based on the analysis of proton number cumulants in heavy-ion collisions. 
Modified from~\cite{Bzdak:2019pkr}.
}
\label{fig:CP}       
\end{figure}

The analysis of (net-)proton number cumulants at different collision energies indicates that experimental data are consistent with non-critical physics such as baryon number conservation and short-range repulsion at $\sqrt{s_{\rm NN}} \gtrsim 20$~GeV. In contrast, the data at lower energies indicate significant excess proton correlations over various non-critical baselines, which require further analysis. 
These observations thus disfavor the existence of QCD CP at small baryon densities, $\mu_B/T < 2-3$, consistent with observations from lattice QCD~\cite{Bazavov:2017dus, Vovchenko:2017gkg, Borsanyi:2021sxv}. 
They also indicate that the CP, if it exists, is located in the baryon-rich matter probed by heavy-ion collisions at intermediate energies~$\sqrt{s_{\rm NN}} \lesssim 20$~GeV, see Fig.~\ref{fig:CP} for the summary of the available CP constraints.
Future efforts in the search for the CP at intermediate collision energies will require improved modeling of CP effects in baryon-rich matter, and new developments in molecular dynamics simulations of critical fluctuations will play a valuable role there.

\section*{Acknowledgments}

This work was supported by the U.S. Department of Energy, 
Office of Science, Office of Nuclear Physics, under contract number 
DE-FG02-00ER41132.



%
%
%

\end{document}